\documentclass{emulateapj}
\usepackage{url,graphicx,amssymb}
\usepackage{subfigure,lscape}

\long\def\symbolfootnote[#1]#2{\begingroup%
\def\thefootnote{\fnsymbol{footnote}}\footnote[#1]{#2}\endgroup}

\newcommand\be{1A 0535+262}

\begin{document}

\shorttitle{\textit{Chandra} grating spectroscopy of \be}
\shortauthors{Reynolds et al.}

\title
{\textit{Chandra} grating spectroscopy of the Be/X-ray binary \be}
\author{Mark T. Reynolds\altaffilmark{1}, Jon M. Miller\altaffilmark{1}}  
\email{markrey@umich.edu}

\altaffiltext{1}{Department of Astronomy, University of Michigan, 500 Church
  Street, Ann Arbor, MI 48109}

\begin{abstract}
We present \textit{Chandra} HETGS spectroscopy of the Be/X-ray binary \be~obtained during the 2009/2010
giant outburst. These are the first CCD grating spectra of this type of system during a giant outburst.
Our spectra reveal a number of lines including a narrow Fe K$\alpha$ emission line with a FWHM of $\rm
\sim 5000~km~s^{-1}$. For the first time, we detect the presence of a highly ionized outflow in a Be/X-ray
binary.  Assuming that the line is He-like Fe XXV, fits with a simple Gaussian imply an outflow velocity
of $\rm \sim 1500~km~s^{-1}$. However, self-consistent photoionization modeling with \textsc{xstar}
suggests that Fe XXIII-XXIV must also contribute. In this case, an outflow velocity of $\rm \sim
3000~km~s^{-1}$ is implied. These results are discussed in the context of the accretion flow in Be-star,
neutron star, and black hole X-ray binaries.
\end{abstract}
 

\maketitle
\section{Introduction}
X-ray binaries are divided into a number of different categories based on their observational
(e.g. X-ray/optical) and physical characteristics . The primary division is made based on the mass of the
secondary star in the binary, with systems for which $\rm M_2 \lesssim 1~M_{\sun}$ being classified as low
mass X-ray binaries (LMXBs), 1 $\rm \gtrsim M_2 \gtrsim 10~M_{\sun}$ classified as intermediate mass X-ray
binaries (IMXBs) and those with secondary masses greater than 10 M$_{\sun}$ classified as high mass X-ray
binaries (HMXBs).  Approximately 300 X-ray binary systems have been detected in the Galaxy, with more than
half of these being LMXBs. Of the approximately 115 known HMXB systems, Be/X-ray binaries comprise the
majority ($\sim$ 70\%; \citealt{c6,c7}).

\begin{figure}
\begin{center}
\includegraphics[width=0.45\textwidth]{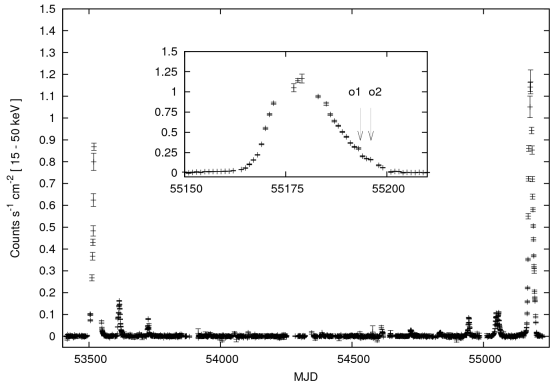}
\caption{\textit{SWIFT} BAT hard X-ray lightcurve for the Be/X-ray binary \be~from 2005 January to 2010
  January. A number of regular outbursts (type-I) are detected in addition to the giant outbursts
  (type-II) in 2005 and 2009. The inset shows a zoom in on the 2009 giant outburst, the lines mark the
  times of our \textit{Chandra} HETGS observations (see text). For reference a
  flux of $\sim$ 0.22 $\rm counts~s^{-1}~cm^{-2}$ is normally measured from the Crab pulsar.}
\label{batlc}
\end{center}
\end{figure}

Be/X-ray binaries comprise of a neutron star primary in a wide eccentric orbit with a Be-star secondary
(no black hole/Be-star binaries have been discovered to date -- see \citealt{c8}). X-ray outbursts in
these systems typically occur during periastron passage of the neutron star close to the circumstellar
excretion disc of the Be-star, e.g. \citet{c9}. However, the X-ray outbursts have been observed to occur
in two distinct flavors, (i) Type-I: Regular outbursts which occur during periastron passage, generally
repeating on the binary orbital period, and (ii) Type-II: Also known as giant outbursts likely arising
from a dramatic expansion of the circumstellar disc. In contrast to type-I outbursts, type-II outbursts do
not have a strict phase dependence and typically last much longer than type-I bursts, with the outburst
time typically being measured in months, e.g. see Fig. \ref{batlc}.

Cyclotron absorption lines have been detected in the X-ray spectra of a number of Be/X-ray binaries
allowing accurate constraints to be placed on the neutron star magnetic field. Typically, magnetic flux
densities in the range $\rm \sim 10^{11} - 10^{13}~G$ are measured, e.g., \citet{c12,c10,c11}. This opens
up the possibility of studying the effect of the neutron star B-field on the accretion flow. In these
HMXBs the magnetic field is much larger than the B-field of the neutron stars in the much older population
of neutron star LMXBs, e.g., a magnetic flux density of $\rm \sim 10^8~G$ has been measured in the
accreting millisecond pulsar SAX J1808.4-3658 \citep{b37}.

\citet{c15} have investigated the timing properties exhibited by sample of 4 Be/X-ray binaries, which
displayed a type II outburst that was observed by \textit{RXTE}. A number of similarities with the well
studied behavior of LMXBs were detected when analyzing the colour--colour (CD) and hardness intensity
diagrams (HID), e.g. \citet{c38,c39}. The main similarities are as follows (i) Similar spectral branches
in the CD/HID as those observed in LMXBs, (ii) smooth motion through the CD/HID, (iii) larger variability
at low intensities, (iv) power spectra that can be described by a small number of Lorentzians, and (v)
flat topped noise at lower frequencies in the horizontal branch that turns into power-law noise in the
diagonal branch. A number of differences were also observed, namely (i) Different patterns in the CD/HID,
e.g. a low intensity soft state, (ii) slower motion through the various patterns (weeks/months as opposed
to hrs/days in LMXBs), (iii) characteristic timescales that are an order of magnitude longer, and (iv) no
apparent correlation between the power spectral parameters (QPOs, RMS) and the mass accretion rate. A
number of features also hint at a B-field dependence of the timing properties, e.g. power-law noise in the
HB branch in V0332+53 but flat topped noise in 4U0115+63 and EXO 2030+375 whose B-fields are 2-3x lower
than that of V0332+53. These observations demonstrate that the accretion flow in the Be/X-ray binaries is
fundamentally similar to that in the more thoroughly studied LMXBs.  

Simultaneous observations at optical/NIR and X-ray wavelengths have shown substantial changes in the
structure of both the accretion and circumstellar discs, with both discs observed to disappear at
times. This changing disc structure is responsible for the varying outburst phases \citep{c17}. The
observed changes in the accretion and circumstellar discs have been successfully explained via the
'truncated disc model' \citep{c9}. Here the circumstellar disc is truncated due to the tidal interaction
with the neutron star. Detailed studies of EXO 2030+375 found that the circumstellar disc is likely
truncated at the 4:1 tidal resonance radius. This is close to the critical radius at periastron. If the
disc extends to, or beyond this radius, matter may escape through the inner Lagrangian point and form an
accretion disc around the neutron star, see \citet{c9} for further details.

\be~is a 103s X-ray pulsar in orbit around a O9.7 -- B0 IIIe companion star \citep{c1}. The $\sim$ 111 day
orbital period is highly eccentric (e $\sim$ 0.47; \citealt{c2}). A total of 7 giant outbursts have been
observed from this system to date -- 1975, 1980, 1983, 1989, 1994, 2005 and 2009, see \citet{c3} for
further details.  In response to the 2009 giant outburst of \be~(\citealt{c4,c5}), we requested and were
granted 20ks of \textit{Chandra} director's discretionary time to obtain a HETGS spectrum during this
outburst. These are the first X-ray grating spectra of a giant outburst from a Be/X-ray binary.

In this paper, we describe observations undertaken with \textit{Chandra} , during the 2009 giant outburst
of the Be/X-ray binary \be. In \S2, we describe the observations. We proceed to analyze the data in \S3,
where we find evidence for the presence of a high velocity highly ionized outflow in this system. In \S4,
these results are discussed and compared to high resolution X-ray spectral observations of a number of
X-ray binaries, and finally our conclusions are presented in \S5.

\section{Observations}
\subsection{\textit{Chandra} HETGS}
A giant outburst from \be~was detected in late 2009 \citep{c4}, the outburst peaked at a 15 -- 50 keV flux
of approximately 6 Crab. Unfortunately, as was the case during the previous 2005 giant outburst, the
source was positioned close to the sun rendering observations impossible over a large part of the
outburst. However, the long duration of a typical type-II outburst from this system (30 -- 60 days)
presented an opportunity to obtain the first high resolution spectrum at X-ray wavelengths of a Be/X-ray
binary in outburst. As such, we requested, and were granted, a \textit{Chandra} DDT observation. \be~was
observed on 2 occasions upon exit from a sun constrained sky position in late December 2009.

In the first observation (obsid: 12066 -- 2009.12.28), the exposure time was $\sim$ 9 ks, while the
exposure time for the second observation (obsid: 12067 -- 2009.12.31) was $\sim$ 11 ks. The high energy
transmission grating spectrometer (HETGS) was used to disperse the incoming photons onto the ACIS-S CCD
array. As the source flux was high, observations were carried out in continuous clocking mode, with a
nominal frame time of $\sim$ 2.8 ms. In this data mode the fast readout times are achieved by collapsing
the data into a single dimension, preserving spectral and timing information at the price of losing one of
the spatial dimensions.

Spectra and lightcurves were extracted from the \textit{Chandra} event lists using \textsc{ciao v4.2}.
Response files were generated using the \textsc{mkrmf} \& \textsc{fullgarf} tasks and first order spectra
from each grating were combined with the \textsc{add\_grating\_orders} task resulting in a single HEG \& MEG
spectrum for each observation. The spectra were binned using the \textsc{grppha} tool to ensure 10 counts
per spectral bin before exporting to \textsc{xspec} for analysis.

\subsection{\textit{RXTE}}
\be~was also observed by \textit{RXTE} during this outburst on an almost daily basis \citep{c40}. Data was
obtained with both the PCA and HEXTE. The \textit{RXTE} observations were quasi-simultaneous with the
\textit{Chandra} observations with pointed observations occurring within a 12hr period of our observations
on each occasion (Obsid: 94323-05-03-03, \*-06; PI: Cabellero). Spectra were extracted in the standard
manner using the relevant \textsc{ftools}. For the PCA, spectra and background files were extracted from
layer-1 of PCU2 alone, giving useful data in the spectral range 3 -- 20 keV. HEXTE spectra were extracted
from cluster-A, while backgrounds were extracted from cluster-B and converted to cluster-A background
files using the \textsc{ftool hextebackest}\footnote{see
  http://heasarc.gsfc.nasa.gov/docs/xte/xhp\_new.html} giving useful data in the 15 -- 100 keV band. As a
number of background features were apparent at higher energies, data above 60 keV was ignored in all
further analysis. Hence, the \textit{RXTE} spectra resulted in a broadband detection of \be~ in the
spectral range from 3 -- 60 keV.

\begin{figure}
\begin{center}
\includegraphics[width=0.35\textwidth,angle=-90]{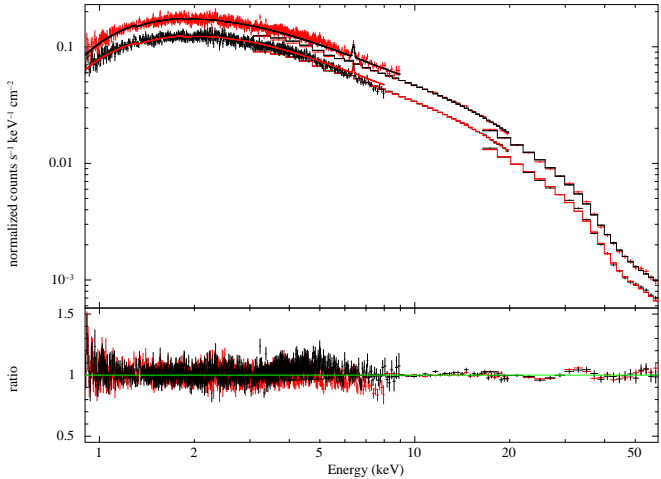}
\caption{Broadband spectra of \be~ as measured by \textit{Chandra} and \textit{RXTE}. The spectra (0.9 --
  60 keV) include data from MEG, HEG, PCA and HEXTE. The top spectrum corresponds to observation 1 (red)
  while the bottom spectrum corresponds to observation 2 (black). The best fit model corresponds to
  \texttt{pha*(cyclabs*(diskbb+bb+cutoffpl))}, see Table \ref{spec_cont} for the best fit parameters.}
\label{bband_spec}
\end{center}
\end{figure}

\section{Analysis \& Results}
The HEG spectra immediately reveal the presence of an emission feature consistent with iron K$\alpha$ at
approximately 6.4 keV in addition to an absorption feature at $\sim$ 6.7 keV. In contrast, outside of a
number of line like residuals in the Si/S region (2.1 -- 2.4 keV), which are likely of an instrumental
nature, there are no obvious spectral features in the MEG spectra. Before we examine in detail the
emission \& absorption lines in the spectrum, we first model the spectral continuum. To this end, we
utilize the \textit{RXTE} data in combination with the \textit{Chandra} data (HEG \& MEG) to constrain the
continuum. All data analysis takes place within \textsc{xspec 12.5.0}. 
 
\subsection{Continuum model}\label{cont_model}
In Fig. \ref{bband_spec}, we plot the broadband \textit{Chandra} and \textit{RXTE} spectrum of \be. The
spectrum was fit with a model consisting of a disk blackbody and a blackbody plus a powerlaw with a
spectral cut-off containing an Cyclotron absorption line modified by interstellar absorption
(\texttt{pha*(cyclabs*(diskbb+bb+cutoffpl))}), as this was consistent with the observed spectrum in the
2005 normal outburst \citep{c3}. The column density was held fixed at a value of $\rm 4.5 \times
10^{21}~cm^{-2}$ consistent with the measured value towards \be~\citep{c21} and with optical measurements,
e.g. \citet{c1}. Allowing the column density to vary results in a best fit value consistent with this
value.  At high energies a broad absorption feature is observed consistent with the expected position of
the fundamental Cyclotron resonance absorption line in this system. The measured energies are consistent
with measurements from the previous 2005 outburst \citep{c10,c3}, and that measured earlier in the 2009
outburst \citep{c40} implying a neutron star magnetic field of $\rm \sim 4 \times 10^{12}~G$.

The resulting best fit model parameters are displayed in Table \ref{spec_cont}.  At higher energies, no
significant change is observed in the position of the Cyclotron lines; however, as the background
dominates above 60 keV, our limited ability to constrain the continuum above the Cyclotron line will
affect our ability to quantify any small changes that may be present.  The simple cut-off power-law model
that we have used to describe the high energy emission provides a good description of the data.  If this
emission arises through Compton up-scattering, then the spectral break may signal the electron temperature
of the hot corona. The low energy spectral break we measure ($\sim$ 20 keV) is consistent with a low
electron temperature if the Comptonizing corona is somewhat optically thick, i.e. $\rm kT_e \sim 10~keV,
\tau \gtrsim 1$.

Assuming a distance to \be~of 2 kpc and a canonical neutron star mass of 1.4 M$_{\sun}$, we measure
unabsorbed 0.9 -- 60 keV luminosities of $\rm L_{X_1} = (1.59 \pm 0.01) \times 10^{37}~erg~s^{-1}$ and $\rm
L_{X_2} = (1.17 \pm 0.01) \times 10^{37}~erg~s^{-1}$, which correspond to Eddington scaled luminosities of
$\sim$ 9\% and 6.6\% for the first and second observation respectively.

\begin{figure}
\begin{center}
\includegraphics[width=0.35\textwidth,angle=-90]{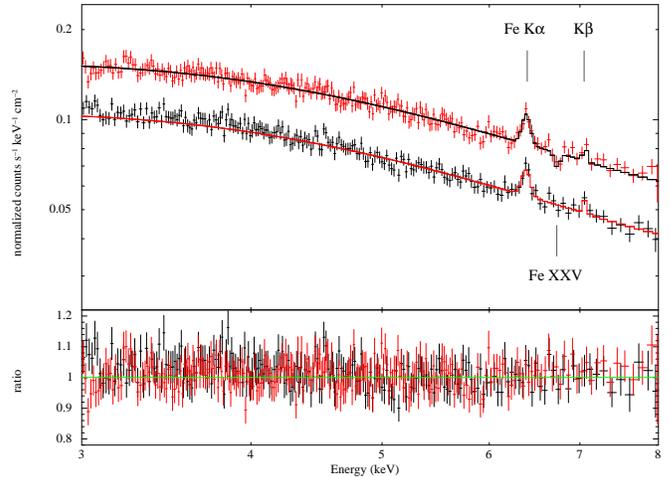}
\caption{\textit{Chandra} HEG spectra of \be~taken at the times indicated in Fig. \ref{batlc}.  Emission
  lines from Fe K$\alpha$ \& K$\beta$ are clearly detected in both observations and are consistent with no
  change within the errors. In contrast absorption consistent with Fe XXV is only detected in the first
  observation. The exposure time was $\sim$ 10 ks in each observation. \textbf{Note:} The data have been
  rebinned for visual clarity.}
\label{heg_spec}
\end{center}
\end{figure}

\begin{figure*}
\begin{center}
\includegraphics[height=0.6\textheight,angle=-90]{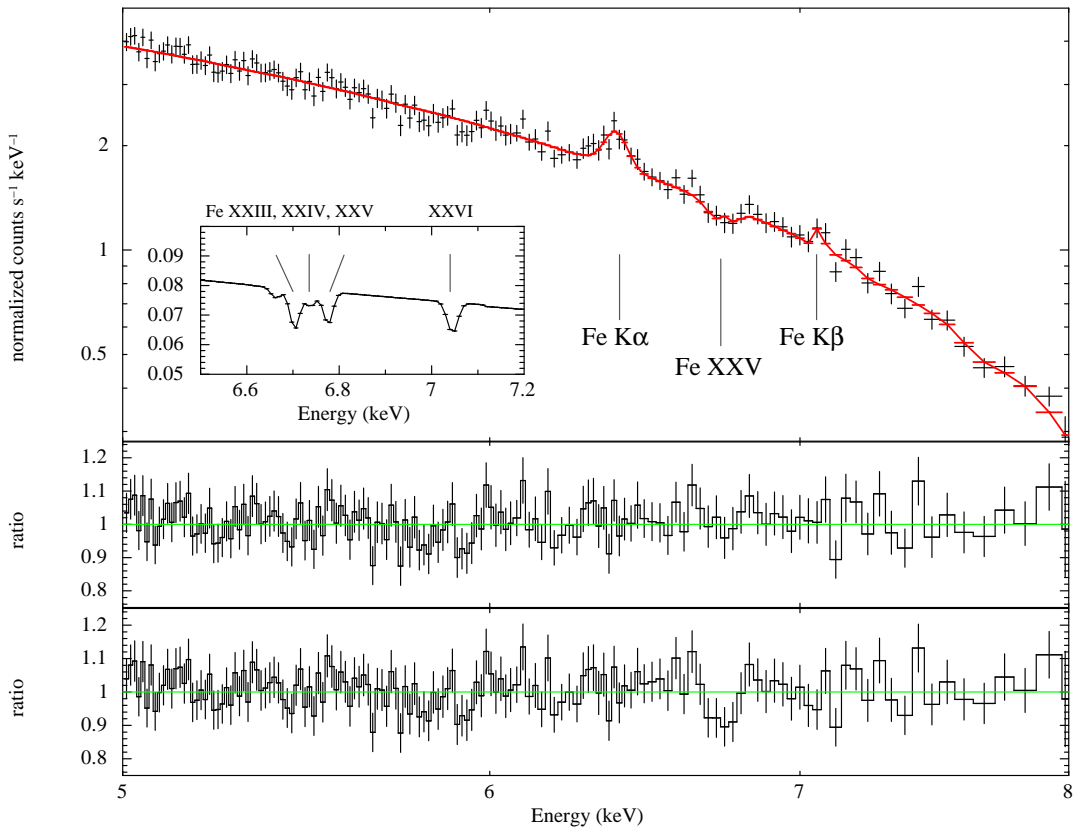}
\caption{\textbf{Main panel:} Zoom in on the 5 -- 8 keV region of the first \textit{Chandra} observation
  with the best fit continuum model. The emission lines have been fit with Gaussians. A single absorption
  line remains in the residuals consistent with an origin from highly ionized Fe XXV (bottom panel). An
  \textsc{xstar} absorption component was then added to the model, the resulting best fit model gives $\rm
  N = 2.0^{+2.0}_{-0.8} \times 10^{21}~cm^{-2},~\xi = 4.0^{+0.3}_{-0.2}~erg~cm~s^{-1}$, and $\rm v_{out} = 3150
  \pm 600~km~s^{-1}$ ($1\sigma$ errors). The absorption line is found to require a contribution from Fe
  XXIII -- XXV, an additional line due to Hydrogen like Fe XXVI is also predicted. \textbf{Inset:} Best
  fit model of the absorption lines in \be, where the y-axis is in units of $\rm
  photons~cm^{-2}~s^{-1}~keV^{-1}$, i.e., the predicted incident spectrum. The normalization of the Fe
  K$\alpha$, K$\beta$ lines have been set to zero in order to emphasize the predicted absorption
  lines. \textbf{Note:} The data in the main panel has not been divided by the detector area, in order to
  illustrate the absence of instrumental features in the Fe K region, and that the data have been rebinned
  for visual clarity.}
\label{abs_line}
\end{center}
\end{figure*}

\subsection{Emission \& absorption lines}\label{simple_model}
Having characterized the continuum, the spectra were then inspected for emission and absorption
features. We concentrate on the HEG spectra as there are no obvious spectral features in the MEG
spectra. A number of emission and absorption features are clearly detected in the iron K region of the HEG
spectrum. In particular, we detect emission consistent with Fe K$\alpha$ and Fe K$\beta$, while a single
absorption line consistent with the presence of highly ionized Fe XXV is also detected in the first
observation alone, see Fig. \ref{heg_spec}.

Basic line parameters were obtained by fitting Gaussians to the detected lines, the parameters of which we
display in Table \ref{spec_lines}. In Fig. \ref{heg_spec}, we display the HEG spectra for each epoch with
the best fit continuum model from \S\ref{cont_model} in addition to a number of Gaussian lines. The iron
K-shell lines are detected in each observation and have the same equivalent widths (EW) within the
errors. The K$\alpha$ line is resolved and hence we measure a line width of $\rm \sim 5000~km~s^{-1}$ in
the first observation while the measured line width is lower in observation 2 at $\rm \sim
4100~km~s^{-1}$, it is nonetheless consistent with the first observation within the error bars (see Table
\ref{spec_lines}).  The observed K$\alpha$ line widths are consistent with velocity broadening in, for
example, an accretion disk. The measured line centroid (6.39$\pm$0.01 keV) supports this, being
inconsistent with the presence of higher ionization states, that would broaden the line, contributing to
the observed Fe K$\alpha$ line, i.e. $>$ Fe X. The measured line fluxes (see Table \ref{spec_lines})
return a Fe K$\beta$/K$\alpha$ flux ratio of 0.21$\pm$0.11 in agreement with the theoretically expected
value within the errors ($\sim$ 0.13, \citealt{c54}).

An absorption line, at an energy consistent with highly ionized Fe XXV is detected in the first
observation alone. We do not observe any absorption from Fe XXVI in either epoch, nor is there evidence
for any additional absorption features in either observation. Assuming the detected absorption line is due
to highly ionized Fe XXV, we measure a significant blue-shift consistent with the presence of an
outflowing wind in the \be~system. The measured line position implies a velocity for the outflowing
material of $\rm v_{out} = 1500\pm1000~km~s^{-1}$.

\subsection{Ionized absorption}\label{xstar}
To provide a more physical description of the highly ionized absorption in the spectrum of \be, we
generated a grid of models using \textsc{xstar} version 2.1kn9, and the "xstar2xspec" facility.  For an
input spectral form, we used the (unabsorbed) best-fit continuum model, extended to cover the 0.1-100.0
keV range.  A covering factor of 0.2 was assumed, consistent with other X-ray binaries (e.g. \citealt{c14}
on GRO J1655-40, \citealt{c18} on GRS 1915+105), and consistent with the absence of any strong ionized
emission lines.  A maximum density of $\rm n = 10^{12}~cm^{-3}$ was assumed, and initially solar
abundances were assumed for all elements. This \textsc{xspec} table model was then added to the continuum
model in \S\ref{cont_model} as an additional absorption component,
i.e. \texttt{pha*(mtable\{xstar.table\}*(continuum))}. The free parameters of this additional absorber are
the column density, $\rm N$, the log of the ionization parameter, $\rm \xi$ and the redshift/blueshift,
$\rm v_{in/out}$.

Fits to the data with this grid of absorption models find the following best-fit parameters: $\rm N =
2.9^{+1.2}_{-1.0}\times 10^{21}~cm^{-2},~\xi = 3.4^{+0.8}_{-0.2}~erg~cm~s^{-1}$, and $\rm v_{out} = 5000
\pm 600~km~s^{-1}$ (uncertainties are $1\sigma$ errors). However, this model predicts H-like Mg and Si
absorption lines that are not observed in the data. Moreover, it appears that the absorption at $\sim$
6.73 keV also contains a contribution from Fe XXIII and Fe XXIV that is comparable to that from the Fe XXV
line.
 
A model where iron is over-abundant by a factor of two provides an improved fit to the data, and does not
predict observable Mg and Si lines.  With such a model, the following absorption parameters are measured:
$\rm N = 2.0^{+2.0}_{-0.8} \times 10^{21}~cm^{-2},~\xi = 4.0^{+0.3}_{-0.2}~erg~cm~s^{-1}$, and $\rm v_{out} =
3150 \pm 600~km~s^{-1}$ (uncertainties are $1\sigma$ errors).  This model provides a fit both to the
entire spectrum and to the ionized Fe K region that is more reasonable. In the Fig. \ref{abs_line} inset,
we display the best fit model containing the \textsc{xstar} absorption component but with the
normalization of the emission lines set to zero for clarity. The incident model spectrum is plotted to
illustrate the contribution from the individual Fe ions. The absorption line is found to be the sum of
absorption from a number of highly ionized Fe ions, i.e. Fe XXIII, Fe XXVI, Fe XXV, see \citet{c55} for
details.

We emphasise, that it is impossible to produce absorption solely due to Fe XXV. In each model where
absorption due to Fe XXV is produced, we also obtain a contribution from the two preceding Fe ionization
stages (XXIII, XXIV). We can estimate the column density in each ion via the standard relation between the
line equivalent width and the curve of growth \citep{c45,c46}. Our best fit model (see
Fig. \ref{abs_line}) implies column densities of $\rm \sim 10^{17}~cm^{-2}$ for Fe XXV and Fe XXIII, while
the Fe XXIV column density is less than a third of this.

The best fit \textsc{xstar} model also predicts absorption due to H-like iron. This absorption line is not
detected in the data due to a combination of the narrow line width, low S/N and relatively low spectral
resolution at this energy ($\sim$ 7.04 keV). This line will contribute to the red wing of the Fe K$\beta$
line, resulting in a larger line EW than measured in \S\ref{simple_model}. In turn this will effect the
measured Fe K$\beta$/K$\alpha$ flux ratio, increasing it from the value measured earlier but nonetheless
remaining consistent with the expected theoretical value within the errors, i.e., $\sim 0.13$
\citet{c54}. However, further observations are required to test for the presence of this Hydrogen like
absorption line.

\begin{table*}
\begin{center}
\caption{Broadband continuum fit parameters}\label{spec_cont}
\begin{tabular}{lccccccccc}
\tableline\\ [-2.0ex]
kT$\rm_{diskbb}$ & norm$\rm_{diskbb}$ & kT$\rm_{bb}$ & norm$\rm_{bb}$ & $\Gamma$ & E$\rm_{cut}$ &
norm$\rm_{cutoffpl}$ & $\rm E_{cyc}$ & $\rm D_{cyc}$ & $\rm W_{cyc}$ \\ [0.5ex]
[ keV ] &  $\rm \times 10^4$ & [ keV ] & $\rm \times 10^{-3}$ & & [ keV ] &  & [ keV ] &  & [ keV ]\\ [0.5ex]
\tableline\tableline\\ [-2.0ex] 
0.15$\pm$0.04        & $5\pm1$ & 1.07$\pm$0.04 & $\rm 7.7\pm0.05$ & 0.54$\pm$0.01 & $\rm
23.0\pm0.2$ & 0.28$\pm0.01$ & 42.85$\pm$0.8 & 0.75$\pm$0.05 & 14.9$\pm$1.3 \\   [0.5ex]
0.17$\pm$0.01 & $3.1\pm0.3$ & $\rm 1.03\pm0.2$  & $\rm 6.3\pm0.03$ &  0.44$\pm$0.1      &
17.6$\pm$0.3 & $\rm 0.183\pm0.004$ & 44.0$\pm$0.5 & 0.57$\pm$0.06 & 8.0$\pm$1.0 \\ [0.5ex] 
\tableline
\end{tabular}
\tablecomments{Best model parameters for the \be~continuum in the spectral range 0.9 -- 60 keV, see
  Fig. \ref{bband_spec}, where the best fit model is \texttt{pha*(cyclabs*(diskbb+bb+cutoffpl))}. All errors
  are quoted at the 90\% confidence level.}
\end{center}
\end{table*}

\section{Discussion}
We present the first high resolution X-ray spectrum of a type-II outburst from the Be/X-ray binary \be.
In the spectra we detect emission lines from Fe K$\alpha$ \& K$\beta$ in two separate observations
separated by 3 days, which do not display any significant evidence for variability between the
observations. In the first epochs observation alone, we find a blue-shifted absorption line consistent
with an origin in highly ionized Fe. Detailed modelling implies an outflow velocity of approximately $\rm
3000~km~s^{-1}$.  We now proceed to discuss the observed spectrum in more detail.

\subsection{The emission line spectrum}
We can use the observed lines to place constraints on the accretion flow geometry.  Firstly, we consider
the observed fluorescence Fe K$\alpha$ emission line. Assuming Keplerian rotation, the line FWHM implies
an emission radius of $\rm r = 7200^{+8600}_{-3500}~km$ in observation 1 and $\rm r =
10800^{+13200}_{-6300}~km$ in the second observation. This emission radius (i.e., $\rm \sim 10^4~km$) is
consistent with the Alfven radius given the known magnetic field in \be~of $\rm \sim 4\times10^{12}~G$,
i.e. $\rm r_A \approx 3.7\times10^{3}~\dot{M}^{-2/7}~km$, where the mass accretion rate is measured in
units of $\rm \sim 10^{-8}~M_{\sun}~yr^{-1}$. Hence, if the Fe fluorescence line originates at the inner
edge of the accretion disc, which is truncated close to the Alfven radius, this implies a mass accretion
rate $\rm \leq 10^{-8}~M_{\sun}~yr^{-1}$. This accretion rate is consistent with the observed X-ray
luminosity during our observation, $\rm L_X \sim 0.09~L_{Edd}$.

We must also ask if the observed fluorescent line strength is consistent with theoretical expectations for
fluorescent line formation. \citet{c24} predict the fluorescent line equivalent width to be $\rm EW \sim
(\Omega/2\pi)*180~eV$. Following \citet{c25}, we find the solid angle subtended by the disc to be $\rm
\Omega \leq 0.25$. The measured EW for the 6.4 keV iron line is $\sim$ 25 eV (see Table \ref{spec_lines}),
implying a solid angle of $\sim$ 0.15. For completeness, we can also ask if the observed line could be
excited on the surface of the mass donor O9.7IIIe star; however, for the known orbital separation at
periastron ($\rm \sim 1\times 10^{13}~cm$, \citealt{c47,c26}) the secondary star will subtend a solid
angle over an order of magnitude less than that of the disc. Similarly the observed line width is
difficult to produce in a stellar wind, for example the typical terminal velocity for the wind from a
B-star is $\rm \lesssim 2000~km~s^{-1}$, this is inconsistent with the measured spectrum. 

Iron fluorescence emission lines have been detected in numerous other neutron star X-ray binaries.     
\citet{c36} have recently completed a systematic analysis of \textit{Chandra} HETGS observations of 41
X-ray binary systems (10 HMXBs \& 31 LMXBs). They find a narrow Fe K$\alpha$ emission line to be present
in all of the HMXBs studied while it is only present in a small fraction of the LMXB sample ($<
10\%$). However, the width of this line is measured to be low, with a typical velocity width of $\rm
\lesssim 800~km~s^{-1}$ and is found to be consistent with a wind like origin. It is clear that the line
width we measure in \be~($\rm v \sim 5000~km~s^{-1}$) is not consistent with having the same origin as the
narrow K$\alpha$ lines detected by \textit{Chandra} in the majority of HMXBs observed to
date. Nonetheless, we note that the line and continuum flux we measure for the Fe K$\alpha$ line are
consistent with the relation $\rm F_{line} \propto F_{cont}^{0.71}$ as measured by \citet{c36} for the
sample of systems they studied. This is to be expected given the small radius implied by the line width
($\rm \sim 10^4~km$). The equivalent width of the Fe K$\alpha$ line is also consistent with the low
significance 'X-ray Baldwin effect' observed in the other HMXBs.

The line we observe in \be~is produced in a similar manner to the relativistically broadened lines that
are observed in many of the neutron star LMXBs, i.e., fluorescent X-ray emission from the surface of the
accretion disc. However, in \be~the observed K$\alpha$ line does not display any evidence of the
characteristic relativistic red wing as expected if the line is formed deep in the gravitational well of
the neutron star, e.g., \citet{c52}. Instead the line we measure is narrow, consistent with a much larger
emission radius ($\rm \sim 10^4~km$, see above).

The emission line spectrum that we observe in the \be~spectrum is fundamentally different to that observed
in the high mass X-ray binary pulsars GX 301-2 \citep{c50} and 4U 1908+075 \citep{c36}. \citet{c50}
observed GX 301-2 for 40 ks with \textit{Chandra} HETGS and detected a fluorescence K$\alpha$ line with an
accompanying Compton shoulder. This line is formed in the dense wind from the B2 supergiant secondary
star. Detailed \textsc{xstar} modelling of this X-ray spectrum found column densities of $\rm N_H \sim
10^{24}~cm^{-2}$. Analysis of a \textit{Chandra} spectrum of 4U 1908+075, reveals a similar Compton broadened
Fe K$\alpha$ line \citep{c36}. In contrast the iron absorption lines detected in \be~are consistent with a
much lower column density in the wind, $\rm N \sim 10^{21}~cm^{-2}$.

\subsection{Wind origin?}
We have measured an absorption component consistent with a significant outflow in this system. Simple
phenomenological modelling favours a outflow velocity of $\rm \sim 1500~km~s^{-1}$ consistent with
absorption due to highly ionized He-like iron, see \S \ref{simple_model}. Detailed modelling with the
\textsc{xstar} plasma photoionization code reveals the absorbing component to consist of a number of
highly ionized Fe ions from XXIII -- XXV at a common blue-shift of $\rm \sim 3000~km~s^{-1}$, see \S
\ref{xstar}. This is the largest wind velocity measured from a Galactic X-ray binary to date.

We can estimate the maximum radius at which the absorption occurs via the measured ionization and column
density from the \textsc{xstar} model, i.e., assuming N=nr, we have $\rm r = L_x/N\xi$, which implies a
radius of $\rm \sim 10^{12}~cm$. This radius is less than the estimated Roche lobe radius of the neutron
star close to periastron ($\rm R_L \sim 3.5 \times 10^{12}~cm$; \citealt{c47,c26}) and implies a disc
origin for the observed absorption component.
  
There are three well studied mechanisms for launching a disc wind (i) radiative -- radiation pressure may
drive a moderately ionized wind \citep{c27} (ii) thermal -- Compton heating of the accretion disc by the
central X-ray source may raise the gas temperature to the escape velocity and drive a thermal wind from
the system \citep{c28}, and (iii) magnetic -- could be powered either by internal magnetic viscosity (MRI)
giving rise to a slow dense wind \citep{c29} or rigid magnetic fields which would allow some material to
be accelerated along magnetic field lines giving rise to a clumpy wind \citep{c30}. Radiative and thermal
winds are expected to be launched from significantly larger radii (e.g., $\rm 10^{10}~cm~vs~10^9~cm$ for a
10$\rm M_{\sun}$ BH) and have a lower density than a magnetic wind, a prediction that may be tested via
detailed modelling of the disc absorption line spectrum as demonstrated by \citealt{c13,c14}. The radius
implied by our modeling favours a radiative/thermal origin for the large outflow velocity measured in \be.

\subsection{Outflows in other X-ray binaries}
\textit{Chandra} observations of GRO J1655-40 \citep{c13,c14} have provided the most revealing insights
into the structure of this accretion disc wind. Detailed studies of the wind absorption line spectrum
favour a magnetic origin with an inner launching radius $\rm \sim 10^9~cm$. The mass flux in the wind was
constrained to be at least 5\% -- 10\% of the total accreted mass and could easily be comparable with the
mass accreted through the disc. Observations of the stellar mass black holes H1743-322 \citep{c22} and 4U
1630-472 \citep{c31} have also displayed evidence for dense winds launched from the inner disc region.
Recent observations of the Galactic microquasar GRS 1915+105 also display blue-shifted absorption features
consistent with an outflow velocity of 150 -- 500 km s$^{-1}$ \citep{c18}. While observations of the Galactic
high mass black binary Cyg X-1 also display evidence for a low velocity outflowing wind \citep{c42}. The
low outflow velocity here is interesting as in this system there is a large scale wind, due to the O-type
secondary star, with an expected terminal wind velocity of $\rm v_{\infty} \leq 2100~km~s^{-1}$.  In all of the
above cases the measured outflow velocity is low $\rm \lesssim 500~km~s^{-1}$.

Similarly, \textit{Chandra} observations of the neutron star LMXB GX 13+1 also measure a low velocity,
highly ionized wind \citep{c32}. Detailed modelling of the observed absorption lines revealed the wind to
be consistent with being radiatively driven. Of the sample of X-ray binaries observed with the HETGS, only
a single system has displayed absorption lines with a velocity comparable to that which we observe
here. \citet{c33,c34} presented spectra of the neutron star binary Cir X-1 \citep{c35}. The Fe XXV
absorption line was found to exhibit a P-Cygni profile consistent with an outflow velocity of $\rm \sim
2000~km~s^{-1}$. In this case it is likely that both thermal and radiation pressure play a significant
role in accelerating the wind to the measured velocity \citep{c34}. In a subsequent \textit{Chandra}
observation of Cir X-1, \citet{c51} unambiguously detect a number of absorption lines which they
identify as highly ionized Fe XXII -- XXV. The detection of these highly ionized iron lines in Cir X-1
would support the results of our \textsc{xstar} modelling which suggest a contribution from multiple Fe
ionization levels to the measured absorption line in \be.
 
The fact that both of the X-ray binaries with the largest observed disc wind velocity (\be, this paper \&
Cir X-1, \citealt{c34}) contain a neutron star primary suggests that the additional hard X-ray emission
originating from the neutron star surface and the disc/neutron star boundary layer is crucial to
accelerate a wind from the disc to the observed velocities ($\rm v_{out} \sim 3000~km~s^{-1}$). Deeper future
observations will no doubt shed light on precise nature of the observed absorption lines and the wind
acceleration mechanism in Be/X-ray binary systems.  Studies of Galactic OB associations often find that Fe
is under-abundant (e.g. \citealt{c19}). The fact that our data and models would appear to require an
over-abundance of iron can be tested via future HETGS observations.

\begin{table*}
\begin{center}
\caption{HEG spectral features}\label{spec_lines}
\begin{tabular}{lccccccccc}
\tableline\\ [-2.0ex]
 & \multicolumn{4}{c}{Observation 1} &  & \multicolumn{4}{c}{Observation 2}\\
 Line & Energy & EW & Flux & $\sigma$ &  & Energy & EW & Flux & $\sigma$\\
 & [ keV ] & [ eV ] & [ $\rm photons~s^{-1}~cm^{-2}$ ] & [ km s$^{-1}$ ] &  & [ keV ] & [ eV ] & [ $\rm
   photons~s^{-1}~cm^{-2}$ ] & [ km s$^{-1}$ ]\\ [0.5ex]
\tableline\tableline\\ [-2.0ex] 
Fe I K$\alpha$ &  6.39$\pm$0.01 & 27 $^{+27}_{-2}$ & $\rm (2.4^{+0.4}_{-0.2}) \times 10^{-3}$ &
5090$^{+1150}_{-1040}$ & | & 6.39$\pm$0.01 & 29 $^{+25}_{-1}$ & $\rm (1.6\pm0.3) \times 10^{-3}$ &
4140$^{+1240}_{-890}$ \\   [0.5ex] 
Fe I K$\beta$ &  7.061$\pm0.015$ & 7$^{+13}_{-3}$ & $\rm (0.5\pm0.2) \times 10^{-3}$ & & | 
& 7.058 & 8$^{+16}_{-6}$ & $\rm (0.4\pm0.2) \times 10^{-3}$ & \\   [0.5ex] 
Fe XXV K$\alpha$ &  6.730$\pm$0.015 & 10$^{+15}_{-3}$ & $\rm (0.8\pm0.3) \times 10^{-3}$ & & | 
& 6.7 & $<$ 3 & $\rm \leq 0.2 \times 10^{-3}$ & \\  [0.5ex] 
\tableline
\end{tabular}
\tablecomments{Atomic lines detected in the \be~HEG spectrum. All errors are quoted at the 1$\sigma$
  confidence level. Note: where lower limits are not given, the lower limit is consistent with
  zero. Likewise parameters without errors were held fixed.}
\end{center}
\end{table*}

\section{Conclusions}
We have presented high resolution X-ray grating spectroscopy of the Be/X-ray binary \be. For the first
time we detect absorption consistent with the presence of a highly ionized outflow in this class of X-ray
binary. When considered together with the previously detected X-ray ionized winds in a variety of other X-ray
binary classes including: black hole X-ray binaries (e.g., high-mass -- Cyg X-1, low-mass -- GRO J1655-40) and
neutron star low-mass X-ray binaries (e.g. Cir X-1, GX 13+1) this would support the emerging picture of a
continued large scale outflow at large accretion rates but in a non-jet form. 

\acknowledgements 
We thank the anonymous referee for his/her careful review and report.  We would like to extend our
gratitude to the CXC director Harvey Tananbaum for his generous allocation of Director's time. We would
also like to thank all of those in the \textit{Chandra} team who contributed to the prompt execution of
this observation. M.R. acknowledges Tim Kallman for his assistance with \textsc{xstar} and Ryan Porter for a
number of useful discussions.  This research made extensive use of the \textit{SIMBAD} database, operated
at CDS, Strasbourg, France and NASA's Astrophysics Data System.




\end{document}